# On the variational framework employing optimal control for biochemical thermodynamics

Adam Moroz* and David Ian Wimpenny

*Corresponding author. E-mail address: adasmaroz@icloud.com

**Abstract**

The maximum energy dissipation principle can be successfully applied to describe a range of basic nonlinear models. The application of the general variational framework (Moroz, 2008, 2009, 2010) has been illustrated for basic linear schemes. The study presented in the paper shows how the kinetic processes at this level of the biomolecular and biophysical phenomena can be effectively described in terms of the maximum energy dissipation principle and its variational formulation. On the basis of this approach, a range of Lagrangians was proposed for basic nonlinear dissipative kinetic models. The results of this study show that the framework is in agreement with nonlinear equilibrium thermodynamics.

## 1. Introduction

The variational approach is generally considered to provide the technical and methodological core of physical formalism unifying various branches of modern physics, including evolutionary physics. Variational formalism has also made considerable progress in thermodynamics [1-4]. Prigogine [5-7] made a significant impact on the development of extreme approaches in thermodynamics. In recent decades, further progress has been made through the work of Sieniutycz in "extended irreversible thermodynamics" [8] by making use of the Prigogine minimum entropy production principle. However, all this progress has overshadowed the understanding of another interesting and alternative principle, the maximum energy dissipation (MED) or maximum entropy production principle [9-25]. The minimum entropy production method has been reviewed on a number of occasions, for example [26-27], including a recent study [28]. These studies have highlighted the limitations of the minimum entropy production approach. Recently considerable progress has been made in the development of maximum energy dissipation principle for variational applications to linear models of dissipative kinetics [29-31] pointing out the direct link to the least action principle.

With respect to the general nonlinear aspect of energetical dissipative processes, some types of nonlinearities are also referred to as synergetics. From an application perspective, the models of synergetic processes, along with the cooperative kinetics, are a source of great interest because they take place across a wide range of fields, including physics, chemistry, biochemistry, biology and engineering. One could expect the variational/Lagrange approach to provide a good methodological and technical basis for the conceptual and technical unification of cooperative phenomena (biology on one hand and physics on another). There is also a clear link to optimal control methods, such as the Pontryagin maximum principle, which have extensive applications in modern biology, biochemistry and biotechnology. Knowledge of the extreme properties of biokinetics is essential in many modern bioengineering fields, including the design and optimization of biochemical system.

In several studies, an optimal control interpretation in a variational approach was employed to formulate the maximum energy dissipation principle for chemical thermodynamics [29-31]. In this paper, we have applied a description of some well-known biochemical models, following our approach [29-31], together with some interpretation of the maximum energy dissipation principle in relation to the least action principle.

## 2. Theory

In previous studies [29-31] a variational formulation for the maximum energy dissipation principle in chemical kinetics was outlined. This employed the positively definite thermodynamic potential and the positively definite dissipation function. We also employed the direct optimal control interpretation of this choice, illustrating it using the OC formulation of classical mechanics, where the control variables appeared as dummy-like variables. Indeed, the classical mechanics variational problem can be reformulated as a dynamic optimal control Lagrange problem [29], for example by the following designation:

$$\dot{q} = u \ , \tag{1}$$

where, the vector of generalised velocities $\dot{q}$ formally becomes a new variable $u$ (e.g. a control variable). In these terms, the corresponding classic mechanical functional for the action becomes

$$S = \int_{t_1}^{t_2} \Lambda(u,q)dt = \int_{t_1}^{t_2} (T(u) - U(q))dt \to extr \ ,$$

$$q(t_1) = q_1 \ , \ q(t_2) = q_2 \ . \tag{2}$$

Here $S$ is the action, $\Lambda$ is the autonomous Lagrange function (Lagrangian), $q$ is vector of generalised coordinates, $U$ is the potential term, $T$ is the kinetic term, $t$ is time, $t_1$ and $t_2$ are the time intervals. The control $u$ should be considered as having no restrictions. In this case, we obtain a classical dynamic Lagrange optimal control problem. The well-known technique to solve this problem (Eqs.(1) and (2)) is based on the Pontryagin maximum principle [32-33]. Then the OC Hamiltonian is

$$H(q,u,p) = -\Lambda(u,q) + p^T u = -T(u) + U(q) + p^T u \tag{3}$$

where $p_i$ is co-state or adjoint vector, and so-called adjoint system of equations for the costate variable

$$\dot{p} = -\frac{\partial H}{\partial q} = \frac{\partial \Lambda}{\partial q} \tag{4}$$

As the control is opened, then for the extremals we can write;

$$\frac{\partial H}{\partial u} \equiv -\frac{\partial \Lambda}{\partial u} + p = 0 \tag{5}$$

and from this equation the extremum of the Hamiltonian can be obtained. In turn, by differentiating this equation, one can obtain

$$\dot{p} = -\frac{d}{dt}\frac{\partial \Lambda}{\partial u} \cdot \tag{6}$$

Now, taking into account (1) that $\dot{q} = u$ and (4) we can obtain the Euler-Lagrange equations. The initial conditions are specified by $q(t_1)=q_1$, and for fixed final time as $q(t_2)=q_2$, therefore the additional requirement of Pontriagin maximum principle of $H=Const$ corresponds to the energy conservation law. The condition $H=0$ arises for a free terminal time approaching infinity under additional transversality assumptions. Conversely, as we have previously highlighted [30], in the case of the optimal control formulation (1-5) of the classical mechanics problem, the interpretation of the Lagrange function $\Lambda$ from (2), as well as the control variable $u$, seems to be cost-like. It is quite difficult to interpret the term $U(q)$ which is negatively defined and responsible for the interactions from the optimal control cost-explicit perspective. On the other hand, one can see that if this term $U(q)$ is positively defined it could simply be interpreted from optimal control in a cost-like manner as the energetical penalty for being in an non-equilibrium state. Moreover, in this case, the character of motion is opposite to the harmonic-like state and is just a relaxation to the equilibrium, this manner is characteristic of all *dissipative* processes (here is idealised that free energy dissipates to heat internally). However, by reformulating in terms of the optimal control and employing a standard optimal control interpretation one can interpret the values and parameters used in relaxative kinetics in a much clearer way.

We can consider the above approach more specifically for chemical kinetics and thermodynamics, as these processes are rather relaxational in nature. In this case the dynamical constraints have more a complex relationship and for a classical mechanics case (1), we can write;

$$\dot{\xi} = f(\xi, u-k), \quad \xi(t_0)=\xi_0 \tag{7}$$

where $\xi$ is the vector of generalised displacements from equilibrium (extent coordinate in chemical thermodynamics), $\dot{\xi}$ is the time-derivative of $\xi$, and $k$ is the vector of the rate constants. Then we can formulate the OC problem and make it more specific to chemical kinetics by choosing the minimization functional in a form;

$$S = \int_{t_0}^{\tau} \Lambda(\xi, u)dt = \int_{t_0}^{\tau} (\Phi(u-k) + \Psi(\xi))dt \to \min \tag{8}$$

following [28] with an open-end, subject to autonomous dynamical system Eq.(7), fixed initial time $t_0$, unspecified final time $\tau$, and a fixed target state $\xi=0$ . Here $S$ is thermodynamic action, $\Lambda$ is thermodynamic Lagrangian, $\Phi = \Phi(u-k)$ is the dissipative function, formally defined an energetic cost (energetic loss for the regulation in case of a metabolic network), for the rate constant deviation from an optimal value $k$ and $\Psi$ is the thermodynamic potential (free energy). We shall consider this OC problem as having no formal restrictions on the control variables $u$ as well as on the state variables $\xi$. Then applying the Pontryagin maximum principle one can construct the optimal control Hamiltonian;

$$H(\xi, u, p) = -\Phi(u-k) - \Psi(\xi) + p^T f \tag{9}$$

and the Pontryagin conditions will be

$$\begin{aligned} &\dot{\xi} = \frac{\partial H}{\partial p} = f(\xi, u); \xi^*(t_0) = \xi_0 \\ &\dot{p} = -\frac{\partial H(\xi^*, u^*, p)}{\partial \xi} = \frac{\partial \Psi}{\partial \xi} - p^T \frac{\partial f}{\partial \xi} \cdot \\ &\frac{\partial H}{\partial u} = -\frac{\partial \Phi}{\partial u} + p^T \frac{\partial f}{\partial u} = 0 \\ &H(\xi^*, u^*, p^*) = \max_u H(\xi^*, u, p^*) \end{aligned} \tag{10}$$

Here the asterisk superscript (*) denotes the optimal trajectory, $p$ is the co-state vector, $p^T$ is transposed vector $p$, final time $\tau$ is free, so no terminal condition is specified. Because of this, at each point of optimal trajectory, the Hamiltonian (9) is equal to zero: $H(\xi^*,u^*,p^*) =0$ [32]. This optimal control outline is more realistic compared to a classic mechanical problem formulation in the OC terms, (1-6).

Reversibly, the problem (7-8) could be formulated as a variational problem when the control variables are singled out. In this case, after resolving equation (7) with respect to $u$ (as it is straight forward for the mechanical case), one can substitute $u$ into (8) and transform the OC problem into the variational form. On this basis we can formulate the pure variational approach to the problem (7-8):

$$S = \int_{t_0}^{\tau} \Lambda(\dot{\xi}, \xi)dt = \int_{t_0}^{\tau} (\Phi(\xi, \dot{\xi}) + \Psi(\xi))dt \to \min, \quad \xi(t_0) = \xi_0, \tag{11}$$

Thus, given that (7) could be resolved with respect to control variable $u$, the problem (7) subject to (8) could be rewritten as a pure variational (in a similar way as for the classic mechanical case). Now for an isolated thermodynamic system, the Lagrange equations become;

$$\frac{\partial^2 \Phi}{\partial \xi \partial \dot{\xi}} \dot{\xi} + \frac{\partial^2 \Phi}{\partial \dot{\xi} \partial \dot{\xi}} \ddot{\xi} - \frac{\partial \Psi}{\partial \xi} - \frac{\partial \Phi}{\partial \xi} = 0 \cdot \tag{12}$$

In terms of the chemical generalised forces $X$ and fluxes $J$, using designations [5, 12]:

$$J \equiv \dot{\xi}; \quad \dot{J} \equiv \ddot{\xi}; \quad X^{\Psi} \equiv -\frac{\partial \Psi}{\partial \xi}; \quad X^{\Phi} \equiv -\frac{\partial \Phi}{\partial \xi}, \tag{13}$$

the Eulere-Lagrange equations Eqs. (12) could be rewritten as

$$A\dot{J} + BJ + X^{\Psi} + X^{\Phi} = 0. \tag{14}$$

where

$$A = \left\| \frac{\partial^2 \Phi}{\partial \dot{\xi}^2} \right\| \text{ and } B = \left\| \frac{\partial^2 \Phi}{\partial \xi \partial \dot{\xi}} \right\|, \tag{15}$$

$X^{\Psi}$ is a generalised force, related to free energy $\Psi$ and $X^{\Phi}$ are generalised forces due to the dependence of the kinetic part (dissipative function) $\Phi$ on the extent coordinate $\xi$. Equation (14) is a differential dynamic relation. However, this equation indicates that, for chemical kinetics, the chemical generalised fluxes are related to the chemical generalised forces by a nonlinear expression. This could be generalised to all non-extended irreversible thermodynamics, where the generalised thermodynamic forces and generalised thermodynamic fluxes can be written in terms of the extent from equilibrium. One can see that in the case when the dissipation function is just a function of only the velocities $\dot{\xi}_i$ (e.g. $\Phi = \Phi(\dot{\xi})$), the coefficients $b_{ij} = 0$ and $a_{ij} = a_{ji}$ (see equation (12). Expressed in a simplified form $\frac{\partial^2 \Phi}{\partial \dot{\xi}_j \partial \dot{\xi}_i} = \frac{\partial^2 \Phi}{\partial \dot{\xi}_i \partial \dot{\xi}_j}$ .

However, because the open-end variational Lagrange problem is formulated, we also need to bear in mind so-called natural boundary conditions when, at the end of relaxation, $\xi_i(\tau) = 0$ :

$$\left( \dot{\xi}^T \frac{\partial \Phi}{\partial \dot{\xi}} - \Phi(\dot{\xi}, \xi) - \Psi(\xi) \right)_{\xi^*, \tau} = H(\xi^*, \dot{\xi}^*) = 0 \cdot$$

This follows the general principle with the additional Pontryagin maximum principle demand for the optimal trajectories (suppose that $\xi^*$ *gives* a local minimum) of the equality of the Hamiltonian to zero $H(\xi^*, p^*) = 0$ in open-end OC problem.

More strict than above free-terminal-time formulation (8) which fixes terminal state at $\xi = 0$ and leaves the terminal time free (however $\xi(t)$ reaches $\xi = 0$ when $t \to \infty$) is the infinite-horizon formulation, when in (11) $\tau = \infty$ . Because $\Lambda(\dot{\xi}, \xi)$ does not depend explicitly on time, the first integral (Beltrami identity) is

$$\Lambda - \dot{\xi} \frac{\partial \Lambda}{\partial \dot{\xi}} = \Phi(\dot{\xi}, \xi) + \Psi(\xi) - \dot{\xi}^T \frac{\partial \Phi}{\partial \dot{\xi}} = C \quad , \tag{16}$$

For a finite-cost infinite-horizon solution relaxing to $\xi = 0$ ($\xi(t) \to 0$ and $\dot{\xi}(t) \to 0$) first integral value (16) is 0, so $C = 0$. The transversality condition for infinite-horizon problem is

$$\lim_{t\to\infty} p^*(t)^T \xi^*(t) = 0. \quad (17)$$

Taking it into account, we could conclude that the part of energy of the system, due to motion in the *M* independent state-variable degrees of freedom, dissipates completely. The equation (16) is, in fact, the energy conservation law for a dissipative (thermodynamic) systems, which states that for the optimal trajectory, the free energy $\Psi$ dissipated is equal to the energy dissipated by the mechanisms formally contained in the dissipative function $\Phi$. The free energy is, from the start, fully dissipated by mechanisms formulated in the dissipation function (in the case of pure variational formulation), or by mechanisms formulated by dynamical constraints, in the OC formulation.

To find the minimum of the action functional (11) which correspond to the Lagrangian second order, the necessary conditions (Legendre conditions) for minimum are:

$$\frac{\partial^2 \Lambda(\xi^*, \dot{\xi}^*)}{\partial \dot{\xi}^2} \geq 0,$$

These are fulfilled in cases similar to the approach [4], when $2\Phi = \dot{\xi}^T R \dot{\xi}$ and $2\Psi = \xi^T L \xi$, for these sufficient conditions the Lagrangian should by jointly convex in $\xi$ and $\dot{\xi}$.

Let us follow the classical example mentioned above [4] when in the vicinity of global equilibrium [29]. Then the Lagrangian can be written in a vector form as

$$\Lambda = \Phi + \Psi = \frac{1}{2}\dot{\xi}^T R \dot{\xi} + \frac{1}{2}\xi^T L \xi \quad (18)$$

where $R$ is the matrix with the elements $r_{ij}$, $L$ is the matrix with elements $l_{ij}$. In a general case, the Euler-Lagrange equations are

$$(R + R^T)\ddot{\xi} = (L + L^T)\xi \text{ or } \ddot{\xi} = R^{-1} L \xi \quad (19)$$

when the matrixes *R* and *L* are symmetric, which also describe the exponential relaxation after applying boundary conditions selecting the stable branch, because of the positive defined scalar matrixes *R* and *L*. The general solution of this equation can be written as a sum:

$$\xi(t) = \xi_0 \cosh(-\sqrt{R^{-1}L}t) + \dot{\xi}_0 (\sqrt{R^{-1}L})^{-1} \sinh(-\sqrt{R^{-1}L}t), \quad (20)$$

where $\xi(t)$ is displacement vector, $\xi_0 = \xi_{t=t1}$, $\dot{\xi}_0 = \dot{\xi}_{t=t1}$ are initial vectors and *R* and *L* are commuting matrixes. The decaying branch is $\xi(t) = \xi_0 \exp(-\sqrt{R^{-1}L}t)$ and this branch requires $\dot{\xi}(t) = \xi_0(-\sqrt{R^{-1}L})$. Then can be written general expression for flux-force relation in terms of generalised thermodynamic fluxes *J* and generalised thermodynamic forces *X*

$$J = \sqrt{R^{-1}L} L^{-1} X, \quad (21)$$

which indicates the relations between the fluxes and forces in a vector form. From this equation, it also follows that for independent dissipative processes in the vicinity of global equilibrium, the generalised thermodynamic fluxes $J_i$ are linear functions of the generalised thermodynamic forces $X_i$.

In a one-dimensional case, the Lagrange function is

$$\Lambda(\xi, \dot{\xi}) = \frac{1}{2} r \dot{\xi}\dot{\xi} + \frac{1}{2} l \xi\xi$$

and the Euler-Lagrange equation will be $r\ddot{\xi} = l\xi$. The formal solution of this well-known equation is a sum of exponents. Taking into account that the problem is stated for an open-end and at $t = \tau$, the extent of the reaction becomes $\xi = 0$, (the open-end problem when $t_2 = \tau$ is not specified). We need to take into account the transversality conditions: the natural boundary conditions should be set because the terminal instantaneous cost-like criterion (the Lagrangian) should be zero at the totally free-end. Then we can find that $\xi = 0$, this in fact repeats the fixed boundary conditions at $t = \infty$, $\xi = 0$. Finally one can obtain the optimal trajectory $\xi^*(t) = \xi_0 \exp(-\sqrt{\frac{l}{r}}t)$, which describes the well-known exponential relaxation to the equilibrium (Fig.1). The calculated numerically trajectory for the co-state variable *p** is shown also in Fig.1, *r=1.0, l=1.0*. Additionally by considering this one-dimensional case in the vicinity of global equilibrium one can obtain the linear relation between correspondent chemical flux *J* and conjugated chemical force *X*: $J = \frac{X}{\sqrt{lr}}$.

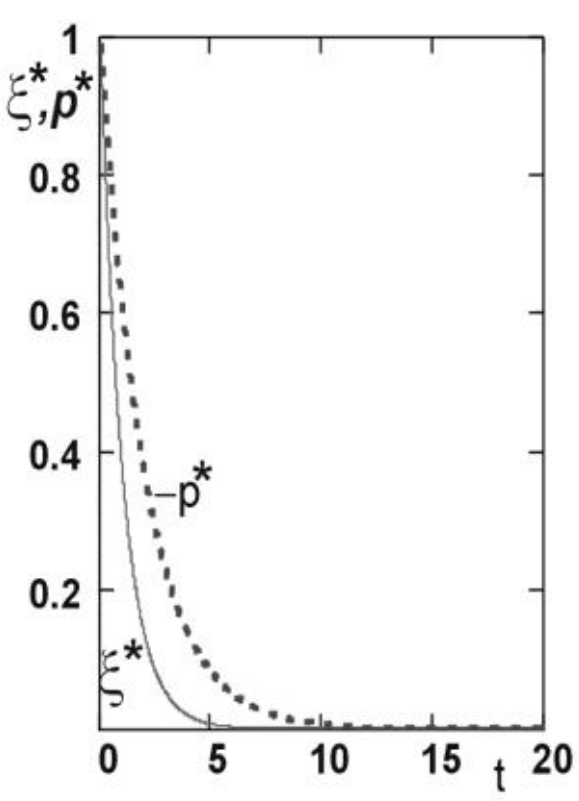

**Fig. 1.** Graphical illustration of one-dimensional quadratic case leading to linear relations and exponential relaxation. The plot of optimal state variable $\xi^*$, and dissipative (thermodynamic) momentum $p^*$ obtained as result of integration of system (26) in one-dimensional case, against time.

Applying the Legendre transform to the thermodynamic Lagrangian from (11) we can obtain the thermodynamic Hamiltonian

$$H(\xi, p) = \dot{\xi}^T(p) p - \Phi(\dot{\xi}(p), \xi) - \Psi(\xi) \quad (22)$$

Then the canonical system could be written as

$$\dot{\xi} = \frac{\partial H}{\partial p}, \dot{p} = -\frac{\partial H}{\partial \xi} = \frac{\partial \Phi}{\partial \xi} + \frac{\partial \Psi}{\partial \xi} \quad (23)$$

In terms of chemical generalised forces *X* and generalised fluxes *J* equations (25) could be rewritten using designations (13), as

$$J_i = \frac{\partial H}{\partial p_i}, \dot{p}_i = \frac{\partial \Phi}{\partial \xi_i} + \frac{\partial \Psi}{\partial \xi_i} \quad (24)$$

We can rewrite the Beltrami first integral (16) for the infinite-horizon optimal trajectory $\xi^*$ as $H(\xi_i^*, p_i^*) = 0$. In Section 3 we will reflect on a further general class for a known non-linear one- dimensional examples.

When matrixes *R* and *L* are scalar symmetric we can find the Hamiltonian using the Legendre transform, when the co-state variables (thermodynamic momenta) *p* are (in a vector form) $p = R\dot{\xi}$, then $\dot{\xi} = R^{-1}p$ and $\dot{\xi}^T = p^T R^{-1}$. Then the Hamiltonian is

$$H(\xi, p) = \frac{1}{2} p^T R^{-1} p - \frac{1}{2}\xi^T L \xi \quad (25)$$

and the canonical system looks very simple because of the symmetry of matrixes *R* and *L* :

$$\dot{\xi} = R^{-1} p$$
$$\dot{p} = L\xi \qquad (26)$$

which could be easy transformed to (21).

Considering the above one-dimensional case, when the dissipation function is just a function on the velocities $\dot{\xi}$ only, e.g. $\Phi = \Phi(\dot{\xi})$ and in the vicinity of global equilibrium, the thermodynamic potential is also a quadratic function $\Psi = \Psi(\xi)$, the thermodynamic momentum is $p = \partial\Lambda/\partial\dot{\xi} = r\dot{\xi}$ and the Hamiltonian will be $H(\xi, p) = \frac{p^2}{2r} - l\frac{\xi^2}{2}$. Then the canonical system will be

$$\dot{\xi}_i = \frac{\partial H}{\partial p} = \frac{p}{r},$$
$$\dot{p}_i = -\frac{\partial H}{\partial \xi} = l\xi$$

or in terms of generalised fluxes and forces: $J_i = \frac{p}{r}, \dot{p}_i = lX$ ·

Also from $H(\xi^*, p^*) = \Phi - \Psi = 0$ for the extremal trajectory one can obtain the linear relation between correspondent chemical flux $J$ and conjugated chemical force $X$.

From equation (22) it is easy to obtain the Hamilton-Jacobi equation [29]. The Hamilton-Jacobi equation could be written as

$$\frac{\partial S}{\partial t} + H(\xi, \frac{\partial S}{\partial \xi}, t) = E \qquad (27)$$

where $S$ is the thermodynamic action in an energetical representation and $H$ is the thermodynamic Hamiltonian that explicitly depends on time. If $H$ does not explicitly depends on $t$, and in case of infinite-horizon problem, $E=0$. In a quadratic form of the dissipative function and potential, the Hamilton-Jacobi equation can be written in a form

$$\frac{1}{2}\left(\frac{\partial S}{\partial \xi}\right)^T R^{-1}\left(\frac{\partial S}{\partial \xi}\right) - \frac{1}{2}\xi^T L\xi = E = 0 \qquad (28)$$

where $\xi$ is the vector of independent generalised displacements.

## 3. Results

The formulated framework (above) was used to study some well-known nonlinear models which are particularly characteristic for nonlinear dissipation. Logistic kinetics describes many cooperative processes in chemical reactions, for example denaturation of polymers, including proteins and DNA, and so on. The results obtained for these models were compared with the classical linear process that gives exponential relaxation, shown in Fig. 1.

Generalising one-dimensional problem for the linear control example to a general nonlinear, we could write the dynamical constraint for the OC problem as

$$\dot{\xi} = uf(\xi) + h(\xi). \qquad (29)$$

That means that the OC problem from (8), in a one-dimensional selection, is subject to an autonomous dynamical system (29). Then this problem could be rewritten as a pure variational problem with the Lagrangian

$$\Lambda(\xi, \dot{\xi}) = \frac{r}{2f(\xi)^2}(\dot{\xi} - h(\xi))^2 + \Psi(\xi) \qquad (30)$$

Then for this case, the Euler-Lagrange system (12) can be written as

$$f\ddot{\xi} - f'_\xi(\dot{\xi}^2 - h^2) = fhh'_\xi + \frac{f^3}{r}\Psi'_\xi \qquad (31)$$

The Beltrami first integral (16) for infinite-horizon problem can be written as

$$\left(\Lambda - \dot{\xi}\frac{\partial\Lambda}{\partial\dot{\xi}}\right)_{\xi^*} = \left(\Psi(\xi) - r\frac{\dot{\xi}^2 - h^2(\xi)}{2f^2(\xi)}\right)_{\xi^*} = 0 \cdot \qquad (32)$$

Giving the costate variable $p$

$$p = \frac{\partial\Lambda}{\partial\dot{\xi}} = \frac{r(\dot{\xi} - h(\xi))}{f^2(\xi)}, \qquad (33)$$

the variational Hamiltonian can be written as

$$H(\xi, p) = \frac{p^2 f^2(\xi)}{2r} + ph(\xi) - \Psi(\xi) \qquad (34)$$

Then the canonical system could be written as

$$\dot{\xi} = \frac{\partial H}{\partial p} = \frac{pf^2(\xi)}{r} + h(\xi)$$
$$\dot{p} = -\frac{\partial H}{\partial \xi} = -\frac{p^2 f(\xi) f'(\xi)}{r} - ph'(\xi) + \Psi'_\xi \qquad (35)$$

In the case of the square form of potential $\Psi(\xi) = l\xi^2/2$ and $h=0$ we can find for the optimal trajectory $\xi^*$ that

$$\dot{\xi}^* = -\sqrt{l/r}\, f(\xi^*)\xi^*. \qquad (36)$$

This first order differential equation can be integrated

$$-\sqrt{\frac{l}{r}}\tau = \int_{\xi_0}^{\xi^*} \frac{dx}{f(x)x}.$$

Also we can find explicitly the optimal costate variable:

$$p^* = -\sqrt{lr}\frac{\xi^*}{f(\xi^*)} \qquad (37)$$

Then the transversality condition following (17) is

$$\lim_{t\to\infty} \sqrt{lr}\frac{\xi^{*2}(t)}{f(\xi^*)} = 0 \cdot$$

If $f(\xi)$ has no singularities at $\xi = 0$, then it hold. From the OC basis it is possible to study more narrowly some functions $f(\xi)$ which can be very useful in applications dealing with basic nonlinear kinetics, including different types of cooperative kinetics leading to the synergetic dissipation of free energy.

### 3.1. *Logistical model*

However, the multiplicative model from (31-39) is too general and further simplifications can be made from the point of view of describing cooperativity. The cooperativity phenomena are well known and have been studied in many fields, including social networks, laser-induced synergetic phenomena or neural network transfer functions, which have logistical form. The variational formulation of logistical/cooperative kinetics could also extend the understanding of this phenomenon from extremal/optimal perspectives.

The Lagrange problem can be reformulated in similar way to the general one-dimensional problem (31-39)

$$S = \int_{t_1}^{\tau}(r\frac{\dot{\xi}^2}{2(1-\xi)^2} + l\frac{\xi^2}{2})dt \to extr \quad , \xi(t_1) = \xi_0 \qquad (38)$$

where $r$ and $l$ are some positively defined constants and free terminal time $\tau$. That means the Lagrangian for the problem is

$$\Lambda(\xi, \dot{\xi}) = \frac{r\dot{\xi}^2}{2(1-\xi)^2} + l\frac{\xi^2}{2} \qquad (39)$$

and the Euler-Lagrange equation is

$$r(1-\xi)\ddot{\xi} + r\dot{\xi}^2 = l(1-\xi)^3\xi \qquad (40)$$

Then using (41) the Hamiltonian will be

$$H(\xi, p) = \frac{p^2}{2r}(1-\xi)^2 - l\frac{\xi^2}{2}. \qquad (41)$$

Now we could write equation for optimal trajectory

$$\dot{\xi}^* = -\sqrt{l/r}(1-\xi^*)\xi^* \qquad (42)$$

which coincides by its form with the well known logistical equation. Its solution is

$$\xi^*(t) = \frac{\exp(-\sqrt{l/r}t)}{C + \exp(-\sqrt{l/r}t)} \qquad (43)$$

and finally in thermodynamic terms the relation between the generalised flux and force is

$$J = \frac{1}{\sqrt{lr}} X + \frac{1}{l\sqrt{lr}} X^2 \cdot \qquad (44)$$

It can be seen that in the vicinity of equilibrium (X<<1), the second term can be ignored and the relationship becomes linear. Then the expression concurs with the expression for ordinary exponential relaxation (linear relations between generalised thermodynamic force and flow), as shown in the first row in Table 1.

The transversality condition $\lim_{t\to\infty} p^*(t)\xi^*(t) = 0$ hold as $p^*\xi^* = \sqrt{lr}\,\frac{\xi^{*2}}{1-\xi^*} \to 0$. Finally along optimal path the two terms in the running cost are equal: $\frac{p^2}{2r}(1-\xi)^2 = l\frac{\xi^2}{2}$.

Fig. 2 graphically illustrates this approach to the application of the logistical kinetics model. Fig. 2A shows the numerical results for different values of the ratio $\sqrt{l/r}$ for the state variable for the logistical model (43). This figure shows the sigmoid type kinetics when the horizontal axis is linear. The dissipative momentum (costate variable) decimal logarithm plot is shown in Fig. 2B, where one can see linear character of the trajectories toward the optimum.

### *3.2. Multiplicative Control Model*

If a one-dimensional free-end terminal case is considered, following (29-37), then the dynamic OC Lagrange problem is

$$S = \int_{t_1}^{\tau} (l\frac{\xi^2}{2} + r\frac{u^2}{2})dt \to extr\ ,\ \xi(t_1) = \xi_0 \qquad (45a)$$

subject to $\dot{\xi} = u\xi$, (45b)

where $k$ and $m$ are some positively defined constants and free terminal time $\tau$ . In the sense of regulation this means that the control is carried out by the rate constant – multiplicatively to the extent variable.

**Table 1**
Summary results for known nonlinear kinetics models (for quadratic approximation of the thermodynamic potential in the vicinity of global equilibrium).

| Relaxation kinetics type | $L(\xi,\dot{\xi})$ | $H(p,\ \xi)$ | $J=J(X)$ |
|---|---|---|---|
| Exponential | $r\frac{\dot{\xi}^2}{2} + l\frac{\xi^2}{2}$ | $\frac{p^2}{2r} - l\frac{\xi^2}{2}$ | $J = \frac{1}{\sqrt{lr}}X$ |
| Logistic | $\frac{r\dot{\xi}^2}{2(1-\xi)^2} + l\frac{\xi^2}{2}$ | $\frac{p^2}{2r}(1-\xi)^2 - l\frac{\xi^2}{2}$ | $J = \frac{1}{\sqrt{lr}}X + \frac{1}{l\sqrt{lr}}X^2$ |
| Logistic-like (multiplicative) | $\frac{r\dot{\xi}^2}{2\xi^2} + l\frac{\xi^2}{2}$ | $\frac{\xi^2}{2}(\frac{p^2}{r} - l)$ | $J = \frac{1}{l\sqrt{lr}}X^2$ |
| Michaelis-Menten like | $\frac{r\dot{\xi}^2(K_m+\xi)^2}{2\xi^2} + l\frac{\xi^2}{2}$ | $\frac{\xi^2}{2}(\frac{p^2}{r(K_m+\xi)^2} - l)$ | $J = \frac{X^2}{l\sqrt{lr}}\frac{1}{K_m + \frac{X}{K_m}}$ |
| Linear | $\frac{r\dot{\xi}^2\xi^2}{2} + l\frac{\xi^2}{2}$ | $-l\frac{\xi^2}{2} + \frac{p^2}{2r\xi^2}$ | $J = \sqrt{\frac{l}{r}}$ |

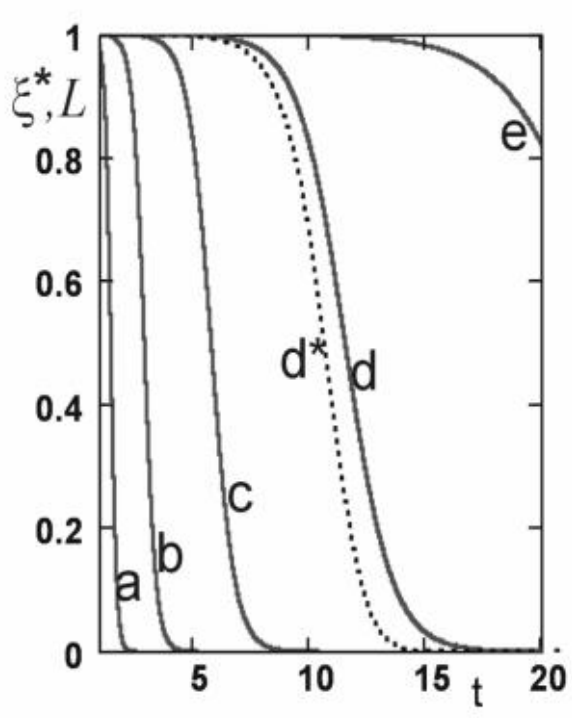

**A**

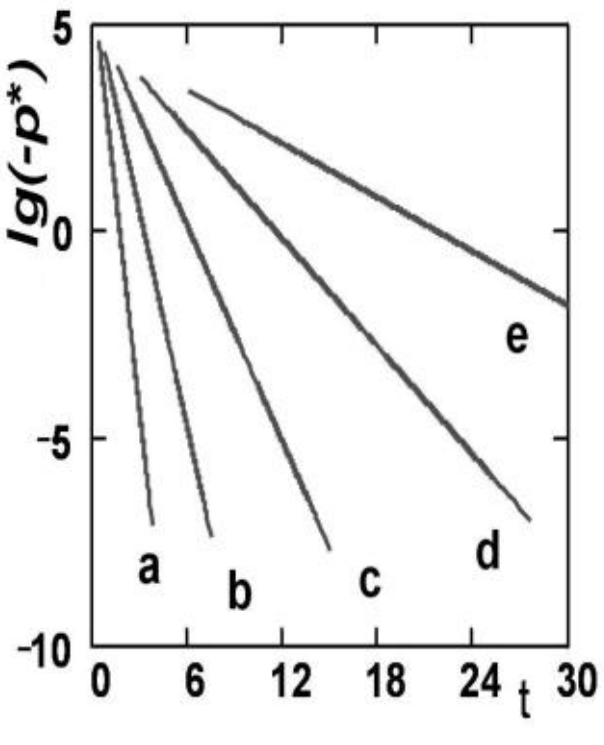

**B**

**Fig. 2**. Illustration of the variational approach to the logistic dissipation type kinetic. (A), the trajectories of state variable $\xi^*$ obtained as numerical solution of correspondent canonical system for different initial conditions: $\xi_{in}=1.0$; a, $p=-10.0$; b, $p=-1.0$; c, $p=-0,1$; d, $p=-0.01$; e, $p=-0.001$, against time, d* - calculated Lagrange function ($L$) corresponding to state variable trajectory d. (B), the plot of the logarithm of numerically calculated dissipative (thermodynamic) momentum $p^*$ against time.

Using a pure variational Lagrange approach to formulate the problem (45), one can write the variational Lagrangian as $\Lambda = r\frac{\dot{\xi}^2}{2\xi^2} + l\frac{\xi^2}{2}$ and obtain the Lagrange equation

$$\ddot{\xi} - \frac{\dot{\xi}^2}{\xi} - \frac{l}{r}\xi^3 = 0 \quad . \qquad (46)$$

Because of the infinite-horizon problem, we need to take into account the Beltrami first integral following (16)

$$\left(\Lambda - \dot{\xi}\frac{\partial\Lambda}{\partial\dot{\xi}}\right)_{\xi^*} = \left(l\frac{\xi^2}{2} - r\frac{\dot{\xi}^2}{2\xi^2}\right)_{\xi^*} = 0 \cdot \qquad (47)$$

The expression inside the brackets actually coincides with the additional demand of the Pontryagin maximum principle for free terminal time for $t = \tau \to \infty$.

On other hand, using a variational Hamiltonian formulation, we could write the Hamiltonian as $H = \frac{p^2\xi^2}{2r} - l\frac{\xi^2}{2}$, then the canonical system will be

$$\dot{\xi} = \frac{\partial H}{\partial p} = \frac{p\xi^2}{r}$$
$$\dot{p} = -\frac{\partial H}{\partial \xi} = -\frac{\xi p^2}{r} + l\xi \qquad (48)$$

This system could be easily obtained by substituting the control variable $u$ for the optimality condition into the equations for the state variable and the costate variable. The system could be reduced to equation $dp/d\xi = \frac{lr}{p\xi} - \frac{p}{\xi}$, where on the right part the equilibrium point $lr - p^2 = 0$ one can find $p_{eq} = \pm\sqrt{lr}$. This result is in the agreement with the equation (47) where same result could be obtained from the additional demand of the Pontryagin maximum principle for the open-end OC problem, when $H(\xi^*,u^*,p^*)=0$ for optimal trajectories. Numerical solutions for system (48) are shown in Fig. 3A, where the set of optimal curves for different $r$ and $l$ combinations can be seen. The relaxation to the equilibrium for extreme state values $\xi^*$ are shown in Fig. 3A for different ratios $l/r$, which are shifted in the direction of the horizontal axes. The plot of the numerically calculated Lagrangian $\Lambda$ is shown in Fig. 3A (for $l=r=1.0$, curve "b*"), where one can see the s-shaped logistical-like curves of the Lagrangian relaxation to zero.

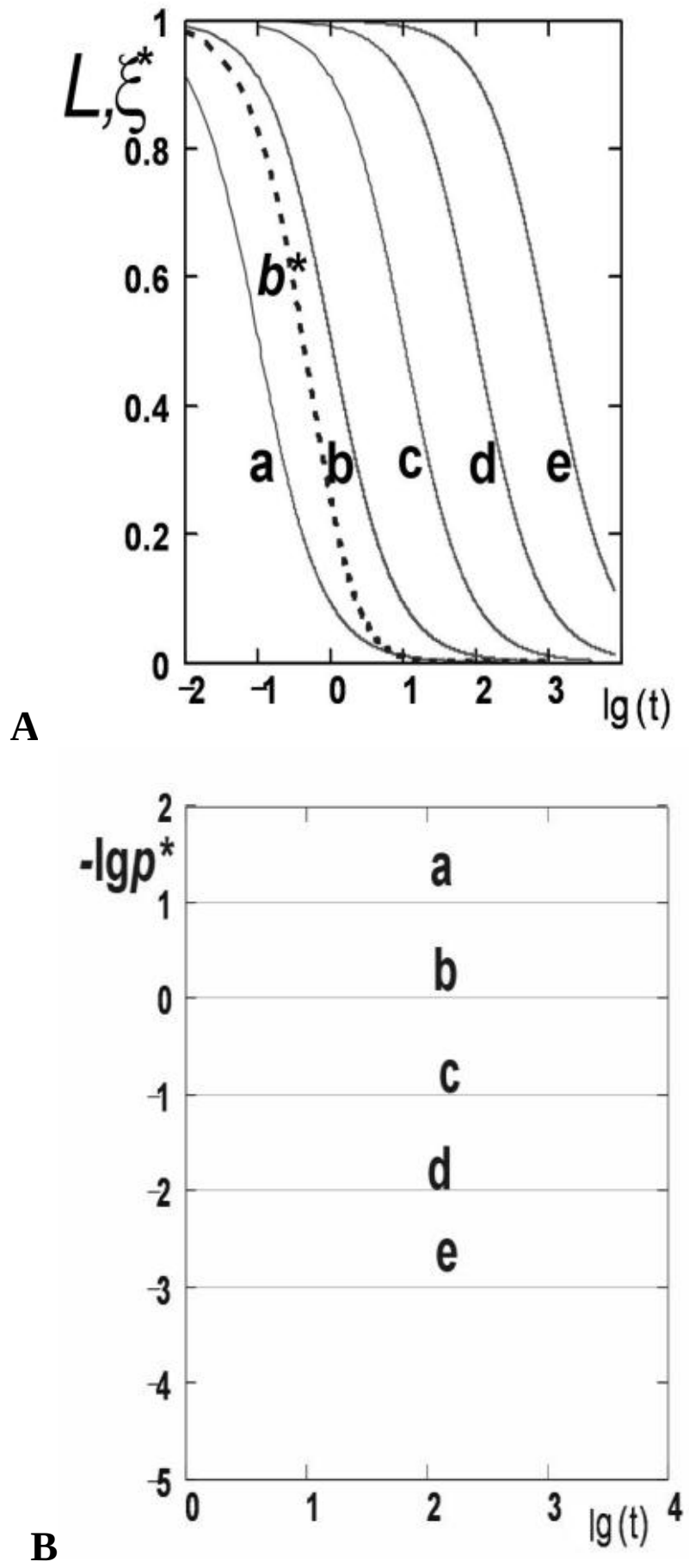

**Fig. 3.** Graphical illustration of the multiplicative model. (A), the plot of state variable $\xi^*$ obtained as numerical solution of system (48), against decimal logarithm of time, $b^*$ - calculated Lagrange function ($L$) corresponding to state variable trajectory b. The plot of the logarithm of numerically calculated dissipative (thermodynamic) momentum $p$ against decimal logarithm of time (B). Initial conditions are: $\xi_{in}=1.0$; a, $p=-10.0$; b, $p=-1.0$; c, $p=-0,1$; d, $p=-0.01$; e, $p=-0.001$.

By comparing Fig.3A and Fig.3B, one can see that co-state variable, which in an optimal state obtains the constant value, influences the parameter $t_{1/2}$ (parameter of the relaxation process when the state variable acquires half of its highest value). This means that the costate variable has real physical meaning in terms of the kinetics of relaxation. In this particular case, it characterises how fast the system described by (48) relaxes.

### *3.3. Michaelis-Menten/Monod kinetics*

The results for the Michaelis-Menten/Monod model are summarized in Fig. 4 for the state variable $\xi^*$ and $p^*$ using (31-39) and function (31) as

$$f(\xi) = \xi \big/ (K_M + \xi). \qquad (49)$$

For this Michaelis-Menten-like model, numerical calculations were performed for different values of the Michaelis constant, $K_M$, responsible for the activation/inhibition of this enzymatic reaction. Fig. 4A shows sigmoid type kinetics with some deviations, curves "a"-"b", to the more complex curves "c"-"f", with a decrease in the dimensionless Michaelis constant $K_M$ from 10 to 0.0001. With a decrease of $K_M$, the energy dissipation rate increases and the area under corresponding curves (action-like value) decreases.

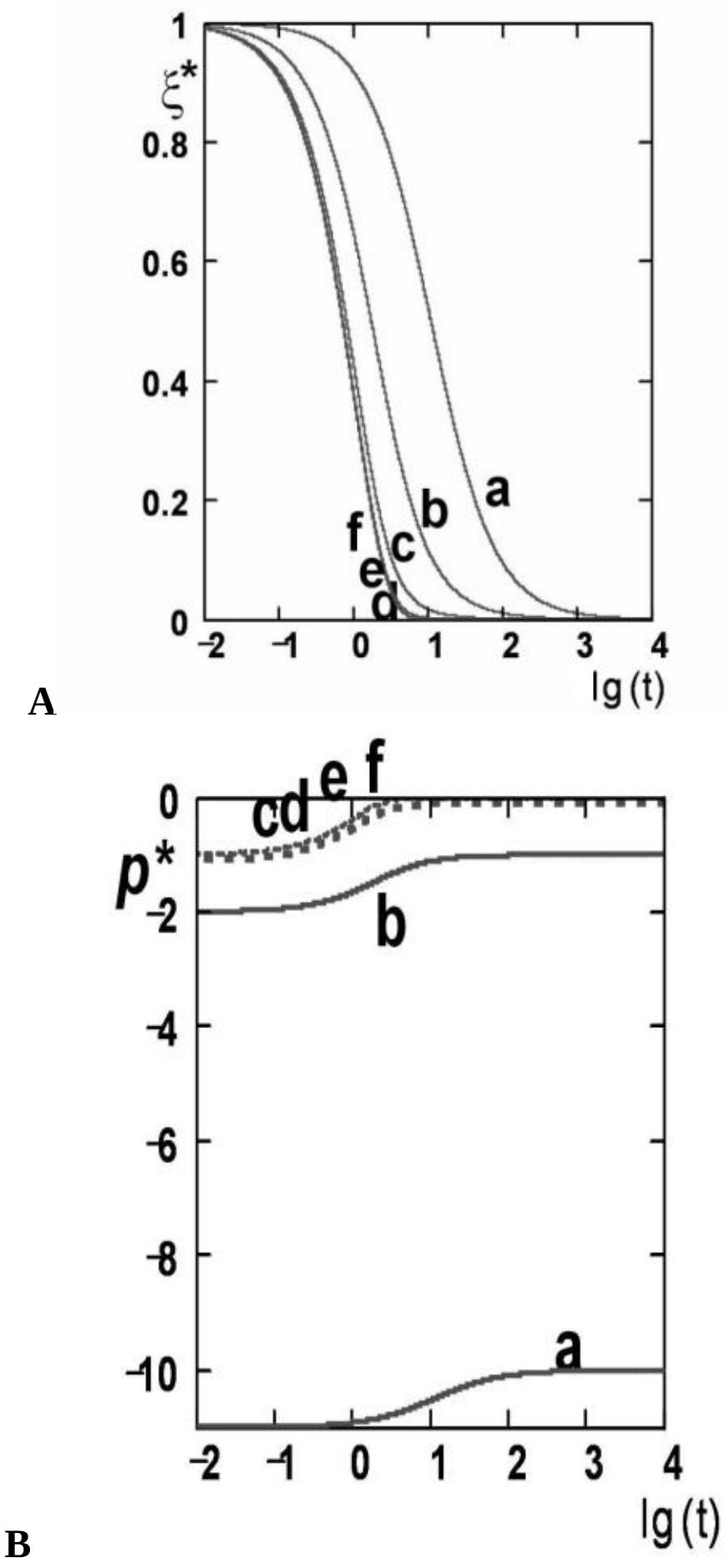

**Fig. 4**. The Michaelis-Menten/Monod type of dissipative kinetics. (A), the plot of state variables $\xi$ obtained as numerical solution for Michaelis-Menten/Monod-like model from table.1 (last column), against decimal logarithm of time. (B), the plot of the logarithm of numerically calculated dissipative (thermodynamic) momentum (costate variable) $p$ against decimal logarithm of time for Michaelis-Menten relaxation. Curve designations: a, $K_M=10.0$; b, $K_M = 1$; c, $K_M = 0.1$; d, $K_M=0.01$; e, $K_M=0.001$; f, $K_M=0.0001$.

There are some similarities in all of the nonlinear models that have been considered. First of all, apart from the linear (exponential relaxation) model, they are cooperative, nonlinear and sigmoid. One can also see the differences in logistic and multiplicative and the Michaelis-Menten-like models. The logistic model is more cooperative, in sense of obtaining equilibrium (compare Fig.2A to Fig.3A or Fig.4A). The logistic co-state variable (thermodynamic momentum) for the logistic model exponentially relaxes, see linear plot in semi-logarithmic vertical coordinate $lg(-p^*)$ for different values of ratio $r/l$, Fig. 2B. For a similar type multiplicative control model (45)-(48), the momentum for optimal trajectory stays at different constant values, Fig. 3B. For the Michaelis-Menten/Monod type of relaxation, the momentum relaxes in a sigmoid manner to equilibrium, Fig. 4B.

Finally, Fig.5 illustrates the kinetics of the relaxation in extent coordinates (Fig. 5A) and dissipated free energy (Fig. 5B)

by linear processes (exponential relaxation) and, as studied above, some nonlinear processes (logistic, quasi-logistic/multiplicative and Michaelis-Menten/Monod-like processes). One can see quite significant differences in the character of the dissipative relaxation of the state variables (general displacements $\xi$). The differences are also in the character of their dissipative momenta (which can be seen from plots of correspondent figures - Fig.1, Fig.2B, Fig.3B, Fig.4B) during their relaxation to the optimal state. Negative values of the dissipative momenta for the real (optimal) relaxation processes mean that the Lagrangian (as well as dissipative potential $\Psi$ and dissipative function $\Phi$) decreases with the time, as one can see from numerically calculated in Fig.2A (curve d*) for logistic model, Fig.3A (curve b*) for multiplicative model, Fig.5B curve b for Michaelis-Menten-like model.

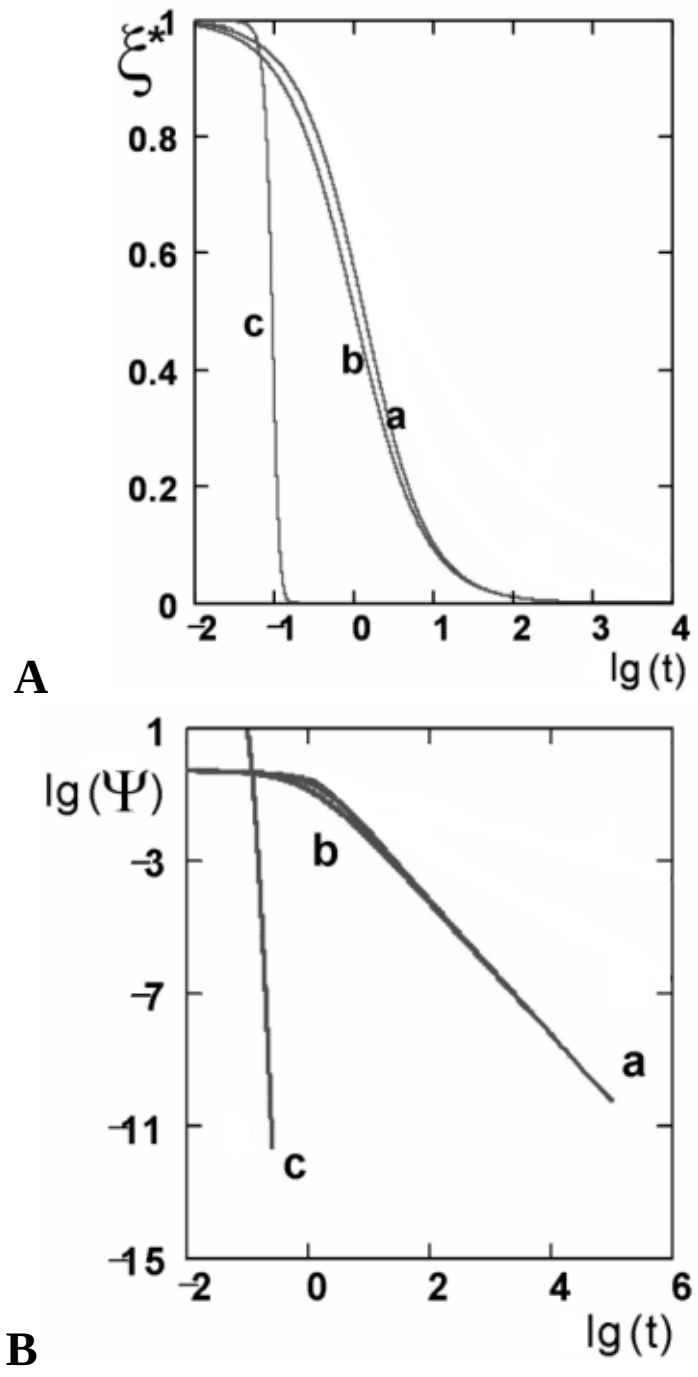

**Fig. 5**. Numerical solutions for different kinetic models: logistical kinetics, multiplicative and Michaelis-Menten (Monod) model. (A), the plot of correspondent state variable $\xi$*, obtained as numerical solution, against decimal logarithm of time. Corresponding to the curves in figure (A), the free energy dissipation plots (B) for its quadratic approximation in double logarithmic scale. Curve designations: a, linear multiplicative function; b, Michaelis-Menten-like ($K_M = 1$); c, logistical ($l$=100). One can see that logistical type has the list square under the curve – that means that in the sense of our interpretation of the principle (coincides in double linear coordinates) is preferable in sense of the control.

Table 1 summarises the results for models considered in the way outlined by (29-37), where generalised fluxes are shown as a functions of generalised forces for different cooperative nonlinear processes. One can see that for the exponential relaxation process, the relationship between flux and force is linear and inversely proportional to the coefficient $r$, which could be interpreted as a resistance. When a nonlinear relaxation process takes place, as in the simplest logistical cooperative process, the relationship between flux and force is much more complex. When included in equation (44), the second part becomes much smaller than the first, the relation then has an exponential form, Table 1, last column. In other cases, when it is not, the second term takes a role and the flux obtains the nonlinear character with respect to dependence on the force.

## 4. Discussion

One can see from the previous section that by applying the MED principle in combination with OC gives the correct description of basic nonlinearities in line with the variational approach in physics. Fig.1, for example, graphically illustrates the exponential relaxation in the linear model. In this model, the state variables $\xi$* decreases exponentially from $\xi_0=1.0$ to zero value, and co-state variable, thermodynamic momentum, also relaxes exponentially from $p_0 = -1.0$ to zero value. Note that the physical sense has the area where $\xi>0$ and $p<0$. The requirement that $p<0$ follows from demand that the functional has to reach a minimum. The physical sense of negatively defined thermodynamic momentum follows from the OC interpretation. In the OC theory, the meaning of the costate variable's (thermodynamic momenta in this particular case) is related to the cost [32]. In an energetical interpretation, negatively defined costs can be interpreted as specific partial energetical loses (free energy loses, its dissipation, when free energy transforms to heat inner to system) and these loses are conjugated to state variables – degrees of freedom of these losses. From the definition of thermodynamic momentum (linear model) $p = \frac{\partial \Lambda}{\partial \dot{\xi}} = r\dot{\xi}$, $r>0$, one can see that in fact in the physical area $\dot{\xi} \le 0$ because of the decrease of $\xi$ with time in the relaxational process.

As one can see (in Fig.2A) the relaxational trajectories for the logistic model are sigmoid in time and the character of the relaxation of thermodynamic momentum is exponential (see Fig.2B). Generally, the application of this approach to the logistic cooperativity model is also in line with physics. In fact, the same character of relaxation can be observed for multiplicative nonlinear dissipation (see Fig. 3A) in a logarithm of time.

The Michaelis-Menten/Monod-like dissipative type kinetics, shown in Fig. 4, matches the above nonlinearities and demonstrates the sigmoid character of relaxation. Fig. 4, also shows the sigmoid character of the relaxation trajectories, Fig.4A as well as for thermodynamic momenta, Fig.4B in logarithm of time. Since the approximation of free energy is quadratic, the relaxation trajectory for free energy will also be sigmoid.

Fig.5A shows a comparison of relaxation, characterized by the state variable $\xi$, for some mechanisms studied. In Fig. 5B, the free energy dissipated in a system by the non-linear methods has a linear character in double logarithmic scale. One can see that, if there is structural variety in this system which allows for different types of non-linear chemical reactions, free energy dissipation could be performed faster as for mechanisms studied. Finally, if the structural variety is very complex, as in the cooperative-like nonlinear processes, the dissipation could have pure logistic nonlinear forms, as was consider in the logistic cooperativity example (see Section 3.1), see Fig. 5A and also Fig. 5B. One can see from Fig. 5B that curve "c" has the fastest dissipation rate (declination coefficient on the graph) and it also corresponds to the least area under the curve "c", Fig. 5A. That indicates that the framework developed above is in line with the least action interpretation, when area under curve of dissipated free energy $\Psi$ (Fig.5B) strives to the minimum. The thermodynamic potential (Gibbs or Helmholtz free energy) achieves its minimum in a way that minimizes the area under the dissipation curve (Fig.5B has an illustrative character, since vertical and horizontal axis are in double logarithmic scale, however qualitatively the pattern/way of will follow the corresponding linear double plot), which at linear double plot in case of free energy [29] has the dimension of the product of energy by time i.e., the physical action. In addition, from an optimal control perspective the non-linear processes have more rigid regulatory characteristics. These are selected in the molecular bioevolutionary process as the efficient regulatory mechanisms to keep important characteristics at the optimal point and with minimum energetical expenses/losses for control.

Now, it seems much clearer that the pure variational approach to dissipative kinetics could be better understood only by using the methodology of the maximum energy dissipation principle and by employing the dynamic optimal control energetical penalty interpretation.

Indeed, for the case when the cost formulation is positively determined, the use of dynamic Optimal Control has certain advantages. Employing its methodology allows us to interpret the dissipative/thermodynamic Lagrange function in a similar way as in the OC theory. So, from the OC perspective, the first term (thermodynamic potential, free energy) is the energetical penalty for being in a nonequilibrium state. The second term (dissipation function) can be interpreted as an energetical expense for performing the dissipation – for the existence of the kinetic mechanisms delivering dissipation. The OC is particularly useful when a search of dynamical constraints is performed, as in equations (29)-(37). After the function $f(\xi)$ is chosen, the kinetic term can then be written as a function of the extent coordinate and its derivative, e.g. $\Phi = \Phi(\xi, \dot{\xi})$. Then the Lagrangian, the Euler-Lagrange equation and costate variable can be written and finally the Hamiltonian. In this way, the optimal control formulation can develop a useful approach from a technical perspective.

Thus, the OC application has a direct influence on the interpretation of the Lagrange function in a 'cost-like' manner. It can be thought as an energetical penalty for being in non-equilibrium and for performing the dissipation. As we can see from Fig. 5B, which comes back to a phenomenological-and-conceptual diagram in [29], the MED principle can specifically outline the thermodynamic formulation of the least action principle for nonlinear kinetics, which is general for all of physics, so for thermodynamics also. Furthermore, the employment of the OC allows for an unambiguous interpretation of the variation approach. Indeed, in mechanics, the physical action and the Lagrange function can be quite difficult to interpret from the optimal control perspective because of the negatively defined potential term in the Lagrangian. In dissipative mechanics and thermodynamics the interpretation becomes much straightforward, because the OC-like appearance of the Lagrangian, e.g. Eq. (11) in the formulation employed above, the optimal control perspective stands more explicitly. The Lagrangian can be interpreted as an instantaneous 'cost-like' function, in other words, the instantaneous energetical penalty function for: a) being in a non-equilibrium state (term $\Psi$) and b) for performing dissipation (term $\Phi$). The Lagrangian is an energetical "penalty" for being in a non-equilibrium state, which should be minimised together with the energetical costs/losses for performing dissipation. The thermodynamic potentials $\Psi$ express the direct energetical "penalty" for being in an unstable/non-equilibrium state and the dissipative function $\Phi$ can be interpreted as an energetical penalty arising from the dissipation kinetics. However, one needs to consider that the Hamiltonian is not related to the energy of the process, as it was in mechanics. Instead it is related to the difference between free energy (potential) and dissipative function, its zero manifests the energy conservation law in the way that free energy in a dissipative process exactly transforms to heat. Optimal trajectory can be expressed by the energy conversation law (free energy dissipated is equal to the free energy dissipated by the kinetical mechanisms in the system), see for example (25). This is why, for the real, extreme trajectory, the Hamiltonian $H(\xi^*, p^*)=0$.

The thermodynamic action, in this sense, could be interpreted as a cumulative energetical penalty, accumulated over the time-period of a dissipative process. Moreover, the least action principle can be interpreted as a physical demand that the dissipation in a thermodynamic/dissipative system should happen as rapidly as possible because the system is penalised for being in a non-equilibrium state. More precisely, should happened in the way that minimises the action under corresponding dissipative curve of the process. Together, the thermodynamic potentials $\Psi$ and dissipative function $\Phi$ construct the instantaneous cost/penalty for being in unstable/non-equilibrium state and for the operation of the dissipation mechanisms in a system. This interpretation could possibly be extended to all thermodynamics, including so-called extended non-equilibrium thermodynamics. The clear optimal control formulation for chemical thermodynamics just makes the interpretation more explicit.

A simple graphical illustration for the free energy dissipation can be developed, which is (with accuracy equal to a constant related to the temperature for an isothermal process) related to thermodynamic entropy production. It can be graphically represented, as shown in Fig. 5B, which further quantifies the diagram from [29]. As the maximum energy dissipation principle requires a maximum free energy dissipation rate, resulting in physical action minimisation under dissipative curve, see [29], it implies that this principle can conceptually correspond to the least action principle. In this case, the least action principle can be treated as a more general principle. It can be also directly extended to the thermodynamic area of the physical processes, in the form of the maximum energy dissipation principle and indirectly in the form of the maximum entropy production principle, when temperature and particularly entropy can be defined/measured adequately for particular thermodynamic system, since for majority dissipative biological processes entropy rather cannot be measured or even calculated exactly. In thermodynamics, the least action principle states that in the field of irreversible transformations of free energy from its various material forms to thermal energy, the rate of these transformations achieves the maximum possible value', as qualitatively illustrated in Figs. 5A and 5B.

All our non-linear examples show a cooperative, synergetic character, however there are a few differences in these examples. The advantage of employing the Michaelis-Menten-like (or Hill) mechanisms/methods of dissipation is that these types of cooperative perturbations can be more effectively controlled. For the Michaelis-Menten mechanism, the method of control is the Michaelis constant, $K_m$. For the Hill mechanism, there are two constants, the dissociation/association constant and the number of binding sites. Consequently, in the sense of regulation, the Michaelis-Menten and Hill mechanisms are controlled types of processes and are, therefore, embedded into more complicated mechanisms of regulation in the hierarchical dissipative processes. The logistical cooperativity looks, in this sense, simpler and provides an overall phenomenological outline.

In our previous studies we interpreted the thermodynamic Hamiltonian as a gain in the energy of the system motion due to its motion in the state variables $\xi_i$ (general displacements/extents). In dissipative processes this gain is zero because the system completely looses (dissipates to heat) energy (more precisely, free energy). From the optimal control perspective it corresponds to additional demand of the Pontryagin maximum principle $H(\xi^*, u^*, p^*)=0$ for an optimal process when the terminal time of the process is indefinite, more exactly, is infinity since the strict problem can be formulated as infinite-horizon problem. Then the first integral, or Boltrami identity can be applied in the form that $H(\xi^*, u^*, p^*)=0$. From the thermodynamic perspective, the equation $H= \Phi - \Psi =0$ states that the energy dissipated in the system is fully dissipated by mechanisms included in the dissipative function $\Phi$; free energy ($\Psi$) dissipated is equal to energy dissipated by $\Phi$. In fact, it can be treated as the energy conservation law in respect to the dissipative kinetics.

It should also be noted that the framework presented in this paper methodologically differs from some other approaches employing the optimal control [8, 34]. These approaches are based on the Prigogine minimum entropy production principle and employ the entropy functional as the action. Therefore, such approaches are at least limited to the steady-state, where the Prigogine theorem applies. Our approach is directly based on the maximum energy dissipation principle, when the thermodynamic Lagrange function has the physical dimension of energy, like in mechanics. The thermodynamic action employed, therefore, has the dimension of energy multiplied by time, as in mechanics and the whole of physics. This approach then allows consistency to be maintained with all physics, including chemical physics. Moreover, this suggests that the maximum energy dissipation principle is just an expression of the least action principle, which is characteristic for all of physics. The least action principle could, in such a sense, be interpreted as a universal principle, according to which, the physical and chemical processes in a system are directed to the extreme elimination of the physical non-equilibrium (expressed by free energy), as far as the structural variety of the system allows.

For basic nonlinear phenomenon, the cooperativity widely taking place in the molecular world can be explained formally and technically in terms of a variational approach, and methodologically in terms of the least action principle (in form of the MED principle as dissipative formulations of the least action principle in thermodynamics). Moreover, the cooperativity at other levels of the biological world (as ligand/enzyme, ligand/receptor binding, macromolecular folding and unfolding, enzyme kinetics, DNA and chromosomes spiral organization, Monod cooperativity in microbiological and biological populations) phenomenologically could also be interpreted as the processes that are capable of accelerating further energy dissipation.

## 5. Conclusion

Within the general framework for maximum energy dissipation principle, where the Lagrangian is a positively defined function of the thermodynamic potentials $\Psi$ and dissipative function $\Phi$, the concrete Lagrangians have been build for a range of non-linear biochemical processes. By interpreting the Lagrangian as a goal function, having an energetical sense from the optimal control perspective, its choice can be explained on the basis that the thermodynamic potential has to be minimized at a possible maximal rate (maximum energy dissipation principle). The Euler-Lagrange equations, which can be written in terms of the generalised thermodynamic forces and generalised thermodynamic fluxes, have been discussed. The first integral following Beltrami, specific for the infinite-horizon model of relaxation process to the global thermodynamic equilibrium, have been discussed and employed to state that the free energy (described by the thermodynamic potential) dissipated equals the energetical expenses of the kinetic dissipative process performed (described by the dissipative function). Therefore, the Hamiltonian is equal to zero on the real trajectory; effectively the form of energy conservation law. The corresponding dissipative Hamilton-Jacobi equation has been discussed for the maximum energy dissipation principle in a general form.

The framework discussed here shows good relevance when employed to the linear and nonlinear dissipative kinetics, well-known in chemical- and bio- applications. In a general one-dimensional case, an important application for these cases has been considered as a basic example. In this example, a number of nonlinear processes characterised by the cooperative energy dissipation were studied. The corresponding Lagrangian and Hamiltonian functions for these processes have been obtained, which give the equations for these processes together with the optimal solutions. The expressions in terms of the generalised thermodynamic forces and generalised thermodynamic fluxes have also been written for these processes. In summary, this study indicates the effectiveness of the proposed framework in describing the basic nonlinear biochemical processes, utilised in free energy dissipation, in terms of maximum energy dissipation. In this respect, the least action principle demands cooperativity phenomena as an organised and quicker method of free energy dissipation. Employing dynamic optimal control is central for the interpretation and choice of a specific dissipative function, when the further pure variational approach is utilized. Use of dynamic optimal control allows to conduct the important interpretations related to the maximum energy dissipation principle. This dissipative interpretation of the least action principle is possible on the basis of dynamic optimal control.

**Acknowledgements**

The authors acknowledge the anonymous reviewers for their valuable comments.

## References

[1] L. Onsager, I. Phys. Rev. 37 (1931) 405-426. II. Phys. Rev. 38 (1931) 2265-2279.
[2] I. Gyarmati, Non-Equilibrium Thermodynamics: Field Theory and Variational Principle, Springer, NewYork, 1970.
[3] M. Biot, Variational Principles in Heat Transfer. Oxford University Press, Oxford, 1970.
[4] B. H. Lavenda, Thermodynamics of Irreversible Processes, MacMillan Press ltd, London, 1978.
[5] I. Prigogine, Introduction use to Non-Equilibrium Thermodynamics, Wiley-Interscience New York, 1962.
[6] P. Glansdorff, I. Prigogine, Thermodynamic Theory of Structure, Stability and fluctuations, Wiley, New York, 1971.
[7] G. Nicolis, I. Prigogine, Self-Organization in Nonequilibrium Systems, Wiley, New York, 1977.
[8] S.Sieniutycz Phys.Reports 326 (2000) 165-258.
[9] H. Ziegler, Principles of Structural Stability, Blaisdell, Waltham, Mass, 1968.
[10] H. Ziegler, ZAMP 21 (1970) 853-863.
[11] H. Ziegler, ZAMP 23 (1972) 553-566.
[12] H. Ziegler ZAMP, 34 (1983) 832-844.
[13] G.W. Paltridge, Q. J. R. Meteorol. Soc. 104 (1978) 927-945.
[14] G.W. Paltridge, Nature 279 (1979) 630-631.
[15] G.W. Paltridge, Q. J. R. Meteorol. Soc. 107 (1981) 531-547.
[16] G.W. Paltridge, Q. J. R. Meteorol. Soc. 127 (2001) 305-313.
[17] R.D. Lorenz, J. Non-Equilib. Thermodyn. 27 (2002) 229-238.
[18] R.D. Lorenz. Inter J Astrobiol 1(1) (2002) 3-13.
[19] H. Ozawa, A. Ohmura, R.D. Lorenz, T. Pujol, Rev. Geophys. 41(4) (2003) 1018.
[20] D. Juretic, P. Županovic, Comp. Bio. Chem. 27 (2003) 541-553.
[21] K. Shizawa, H.M. Zbib, Int. J. Plasticity 15 (1999) 899-938
[22] A. Kleidon, R.D. Lorenz (Eds.), Non-equilibrium Thermodynamics and the Production of Entropy in Life, Earth, and Beyond, Springer, Heidelberg, 2004.
[23] A. Kleidon, K. Fraedrich, T. Kunz, F. Lunkeit, Geophys. Res. Lett. 30 (23) (2003) 2223-2233.
[24] R.C. Dewar, J. Phys. A: Math. Gen. 36 (2003) 631-641.
[25] R. C. Dewar, J. Phys. A: Math. Gen. 38 (2005) L371-L381.
[26] P. Zupanovic, D. Juretic and S. Botric, FIZIKA A 14 (2005) 89-96.
[27] R.C. Dewar, D. Juretic and P. Zupanovic, Chem. Phys. Lett. 430 (2006) 177-182.
[28] L.M. Martyushev, V.D. Seleznev, Physics Reports 426 (2006) 1-45.
[29] A. Moroz, Chem. Phys. Lett. 457 (2008) 448-452.
[30] A. Moroz, J. Phys. Chem. B 113 (2009) 8086–8090.
[31] A. Moroz, Phys. Lett. A 374 (2010) 2005-2010.
[32] I.M. Gelfand, S.V. Fomin, Calculus of Variation, Prentice-Hall, Englewood Cliffs, NJ, 1963.
[33] L.S. Pontryagin, V.G. Boltyanskii, R.V. Gamkrelidze, E.F. Mischenko, The mathematical Theory of Optimal Processes, Interscience, New York, 1962.
[34] S. Sieniutycz. Catalysis Today 66 (2001) 453-460.